%% file: HC3N_revise.tex
\documentclass{aa}
\usepackage{natbib}
\usepackage{graphicx}
\usepackage{txfonts}
\usepackage{aalongtable}
\bibpunct{(}{)}{;}{a}{}{,}
\newcommand{\cc}{$\mathrm{HC_3N}$\,}
\newcommand{\hh}{$\rm H_2$\,}

\begin{document}
\title{Rotational Excitation of HC$_3$N by H$_2$ and He at low
    temperatures}
%
%
 \author{ M. Wernli \and L. Wiesenfeld \and A. Faure
   \and P. Valiron }
\institute{Laboratoire d'Astrophysique de l'Observatoire de
    Grenoble, UMR 5571 CNRS/UJF,
    Universit\'e Joseph-Fourier, Bo\^\i te postale 53, 38041
    Grenoble cedex 09, France}
  \date{Received \today / Accepted }
\abstract {} {Rates for rotational excitation of \cc by collisions
with He atoms and H$_2$ molecules are computed for kinetic
temperatures in the range 5$-$20~K and 5$-$100~K, respectively.}
{These rates are obtained from extensive quantum and
quasi-classical calculations using new accurate potential energy
surfaces (PES). The \cc -- He PES is in excellent agreement with
the recent literature. The \cc -- H$_2$ angular dependence is
approximated using 5 independent H$_2$ orientations. An accurate
angular expansion of both PES suitable for low energy scattering
is achieved despite the severe steric hindrance effects by the \cc
rod. } {The rod-like symmetry of the PES strongly favours even
$\Delta J$ transfers and efficiently drives large $\Delta J$
transfers. Despite the large dipole moment of {\cc}, rates
involving ortho-H$_2$ are very similar to those involving
para-H$_2$, because of the predominance of the geometry effects.
Excepted for the even $\Delta J$ propensity rule, quasi classical
calculations are in excellent agreement with close coupling
quantum calculations. As a first application, we present a simple
steady-state population model that shows population inversions for
the lowest HC$_3$N levels at H$_2$ densities in the range
10$^4-$10$^6$~cm$^{-3}$. } {The \cc molecule is large enough to
present an original collisional behaviour where steric hindrance
effects hide the details of the interaction. This finding,
combined with the fair accuracy of quasi classical rate
calculations, is promising in view of collisional studies of
larger molecules.}


\titlerunning{Cyanoacetylene rotational excitation}
\maketitle


\section{Introduction}

Cyanopolyyne molecules, with general formula HC$_{2n+1}$N, $n\ge
1$, have been detected in a great variety of astronomical
environments and belong to the most abundant species in cold and
dense interstellar clouds \citep{bell97}. One of these,
HC$_{11}$N, is currently the largest unambiguously detected
interstellar molecule \citep{bell85}. The simplest one, \cc
(cyanoacetylene), is the most abundant of the family. In addition
to interstellar clouds, \cc has been observed in circumstellar
envelopes \citep{pepe04}, in Saturn satellite Titan
\citep{kunde81}, in comets \citep{bockelee00} and in extragalactic
sources \citep{mauersberger90}. Furthermore, \cc has been detected
both in the ground level and in excited vibrational levels, thanks
to the presence of low-lying bending modes
\citep[e.g.][]{wyrowski03}. Owing to a low rotational constant and
a large dipole moment, cyanoacetylene lines are thus observable
over a wide range of excitation energies and \cc is therefore
considered as a very good probe of physical conditions in many
environments.

Radiative transfer models for the interpretation of observed \cc
spectra require the knowledge of collisional excitation rates
participating to line formation. To the best of our knowledge, the
only available collisional rates are those of \citet{green78} for
the rotational excitation of HC$_3$N by He below 100~K. In cold
and dense clouds, however, the most abundant colliding partner is
H$_2$. In such environments, para-\hh is only populated in the
$J=0$ level and may be treated as a spherical body.
\citet{green78} and \citet{dickinson82} postulated that the
collisional cross-sections with para-\hh $(J=0)$ are similar to
those with He (assuming thus an identical interaction and
insensitivity of the scattering to the reduced mass). As a result,
rates for excitation by para-\hh were estimated by scaling the
rates for excitation by He while rates involving ortho-\hh were
not considered.

In the present study, we have computed new rate coefficients for
rotational excitation of \cc by He, para-\hh ($J=0$) and
ortho-\hh($J=1$), in the temperature range 5$-$20~K for He and
5$-$100~K for H$_2$. A comparison between the different partners is
presented and the collisional selection rules are investigated in
detail. The next section describes details of the PES calculations.
The cross-section and rate calculations are
presented in Section~\ref{sec:cross}. A discussion and a first
application of these rates is given in
Section~\ref{sec:disc}. Conclusions are drawn in Section~5. The
following units are used throughout except otherwise stated: bond
lengths and distances in Bohr; angles in degrees; energies in
cm$^{-1}$; and cross-sections in $\AA^2$.


\section{Potential energy surfaces}\label{sec:pot}

Two accurate interatomic potential energy surfaces (PES) have recently
been calculated in our group, for the interaction of \cc with He and
H$_2$. Both surfaces involved the same geometrical setup and similar
\textit{ab initio} accuracy. An outline of those PES is given below,
while a detailed presentation will be published in a forthcoming
article.

In the present work, we focus on low-temperature collision rates, well
below the threshold for the excitation of the lower bending mode
$\nu_7$ at 223 cm$^{-1}$. The collision partners may thus safely be
approximated to be rigid, in order to keep the number of degrees of
freedom as small as possible. For small van der Waals complexes,
previous studies have suggested \citep{jeziorska00,jankowski05} that
properly averaged molecular geometries provide a better description of
experimental data than equilibrium geometries ($r_e$ geometries). For
the $\rm H_2O$ -- \hh system, geometries averaged over ground-state
vibrational wave-functions ($r_0$ geometry) were shown to provide an
optimal approximation of the effective interaction \citep{faure05,
wernlithese}.

Accordingly, we used the \hh bond separation $r_{\rm HH}= 1.44876$
Bohr obtained by averaging over the ground-state vibrational
wave-function, similarly to previous calculations
\citep{hodges04,faure05,wernli06}. For \cc, as vibrational
wave-functions are not readily available from the literature, we
resorted to experimental geometries deduced from the rotational
spectrum of \cc and its isotopologues (\citealt{thor00}; see also
Table~5.8 in \citealt{gordy}). The resulting bond separations are the
following:
$r_{\mathrm{HC_1}}= 1.998385$;\ $r_{\mathrm{C_1C_2}}=2.276364$;\
$r_{\mathrm{C_2C_3}}= 2.606688$;\ $r_{\mathrm{C_3N}}= 2.189625$, and
should be close to vibrationally averaged values.

For the \cc -- He collision, only two coordinates are needed to fully
determine the overall geometry. Let $\vec{R}$ be the vector between
the center of mass of \cc and He. The two coordinates are the distance
$R=|\vec{R}|$ and the angle $\theta_1$ between the \cc rod and the
vector \textbf{\textit{R}}. In our conventions, $\theta_1 = 0$
corresponds to an approach towards the H end of the \cc rod. For the
collision with H$_2$, two more angles have to be added, $\theta_2$ and
$\phi$, that respectively orient the \hh molecule in the
rod-\textbf{\textit{R}} plane and out of the plane. The \cc -- He PES
has thus two degrees of freedom, the \cc -- \hh four degrees of
freedom.

As we aim to solve close coupling equations for the scattering, we
need ultimately to expand the PES function $V$ over a suitable angular
expansion for any intermolecular distance $R$. In the simpler case of
the \cc -- He system, this expansion is in the form:
\begin{equation}\label{eq:pot}
     V_{}(R,\theta_1) = \sum_{l_1}
     v_{l_1}(R)\,P_{l_1}(\cos\theta_1)\quad ,
\end{equation}
where $P_{l_1}(\cos\theta_1)$ is a Legendre polynomial and
$v_{l_1}(R)$ are the radial coefficients.

For the \cc -- \hh system, the expansion becomes:
\begin{equation}\label{eq:pot2}
  V(R,\theta_1, \theta_2, \phi) = \sum_{l_1 l_2 l} v_{l_1 l_2 l}(R)
    s_{l_1 l_2 l}(\theta_1, \theta_2, \phi),
  \end{equation}
where the basis functions $s_{l_1 l_2 l}$ are products of spherical
harmonics and are expressed in Eq.~(A9) of \citet{green75}. Two new
indices $l_2$ and $l$ are thus needed, associated respectively with
the rotational angular momentum of \hh and the total orbital angular
momentum, see also eq. (A2) and (A5) of \citet{green75}.

\begin{figure}
  \centering \includegraphics[height=0.55\textheight]{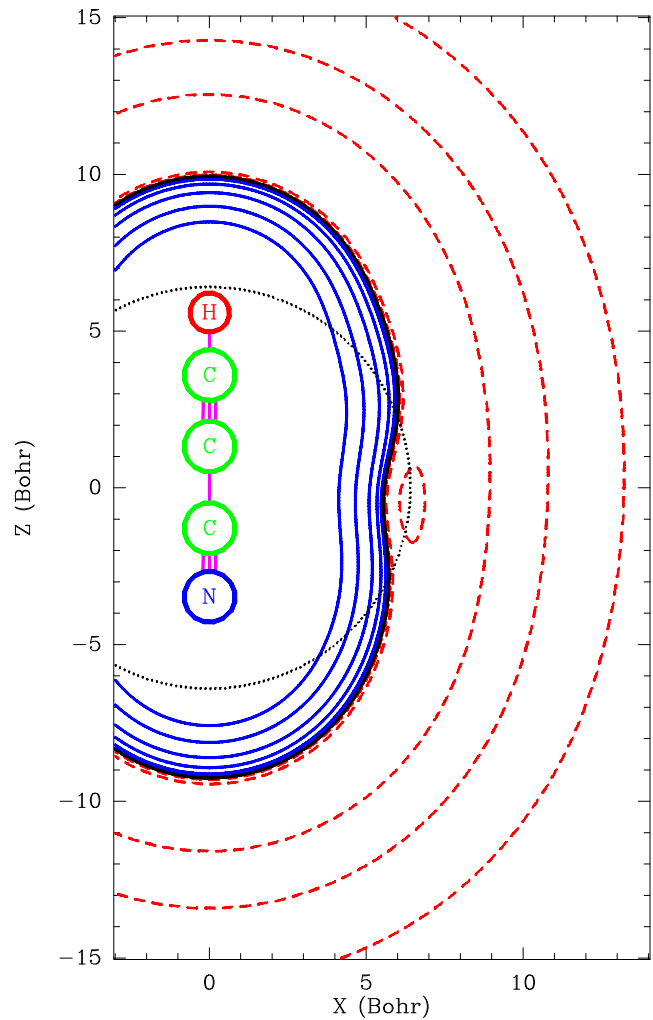}
  \caption{The \cc -- para-\hh PES. The \cc molecule is shown at
    scale. Equipotentials (in $\rm cm^{-1}$)~: in dashed red, -100,
    -30 -10, -3; in solid black, 0; in blue, 10, 30, 100, 300, 1000,
    3000. The dotted circle centered at the \cc center of mass with
    radius $R=6.41$ Bohr illustrates the angular steric hindrance
    problem occurring when the collider rotates from the vicinity of
    the minimum towards the \cc rod.  }
  \label{fig:PES}
\end{figure}

Because the Legendre polynomials form a complete set, such
expansions should always be possible. However, \citet{chapman77}
failed to converge above expansion (\ref{eq:pot}) due to the
steric hindrance of He by the impenetrable \cc rod, and
\citet{green78} abandoned quantum calculations, resorting to quasi
classical trajectories (QCT) studies. Similar difficulties arise
for the interaction with H$_2$.  Actually, as can be seen on
figure~\ref{fig:PES} for small $R$ values, the interaction is
moderate or possibly weakly attractive for $\theta_1 \sim
90^{\circ}$ and is extremely repulsive or undefined for $\theta_1
\sim 0, 180^{\circ}$, leading to singularities in the angular
expansion and severe Gibbs oscillations in the numerical fit of
the PES over Legendre expansions.

Accordingly, we resorted to a cautious sampling strategy for the PES,
building a spline interpolation in a first step, and postponing the
troublesome angular Legendre expansion to a second step. All details
will be published elsewhere. Let us summarize this first step for He,
then for H$_2$.

For the \cc -- He PES, we selected an irregular grid in the
$\left\{R,\theta_1\right\}$ coordinates. The first order derivatives
of the angular spline were forced to zero for $\theta_1=0,180^{\circ}$
in order to comply with the PES symmetries. For each distance, angles
were added until a smooth convergence of the angular spline fit was
achieved, resulting to typical angular steps between 2 and
15$^{\circ}$. Then, distances were added until a smooth bicubic spline
fit was obtained, amounting to 38 distances in the range 2.75 -- 25
Bohr and a total of 644 geometries. The resulting PES is perfectly
suited to run quasi classical trajectories.


We used a similar strategy to describe the interaction with H$_2$,
while minimizing the number of calculations. We selected a few
$\left\{\theta_2,\phi\right\}$ orientation sets, bearing in mind
that the dependence of the final PES with the orientation of \hh
is weak. In terms of spherical harmonics, the PES depends only on
$Y_{l_2m_2}(\theta_2,\phi)$, with $l_2=0,2,4,\ldots$ and
$m_2=0,1,2,\ldots$, $|m_2|\leq l_2$. Terms in $Y_{l_2m_2}$ and
$Y_{l_2 -m_2}$ are equal by symmetry. Previous studies
\citep{faure05a,wernli06} have shown that terms with $l_2> 2$ are
small, and we consequently truncated the $Y_{l_2m_2}$ series to
$l_2\leq 2$. Hence, only four basis functions remain for the
orientation of \hh: $Y_{00},Y_{20},Y_{21}$ and $Y_{22}$.

Under this assumption, the whole \cc -- \hh surface can be obtained
knowing its value for four sets of $\left\{\theta_2, \phi\right\}$
angles at each value of $R$. We selected actually five sets, having
thus an over-determined system allowing for the monitoring of the
accuracy of the $l_2$ truncation. Consequently, we determined five
independent PES, each being constructed similarly to the \cc -- He one
as a bicubic spline fit over an irregular grid in $\left\{R,
\theta_1\right\}$ coordinates. The angular mesh is slightly denser
than for the \cc -- He PES for small $R$ distances to account for more
severe steric hindrance effects involving \hh. In total, we computed
3420 $\left\{R, \theta_1, \theta_2, \phi\right\}$ geometries. Finally,
the \cc -- H$_2$ interaction can be readily reconstructed from these
five PES by expressing its analytical dependence over
$\left\{\theta_2, \phi\right\}$ \citep{wernlithese}.

For each value of the intermolecular geometry
$\left\{R,\theta_1\right\}$ or $\left\{R, \theta_1, \theta_2,
\phi\right\}$, the intermolecular potential energy is calculated at
the conventional CCSD(T) level of theory, including the usual
counterpoise correction of the Basis Set Superposition Error
\citep{jansen69,boys70}. We used augmented correlation-consistent
atomic sets of triple zeta quality (Dunning's aug-cc-pVTZ) to describe
the \cc rod. In order to avoid any possible steric hindrance problems
at the basis set level, we did not use bond functions and instead
chose larger Dunning's aug-cc-pV5Z and aug-cc-pVQZ basis set to
better describe the polarizable (He, H$_2$) targets, respectively. All
calculations employed the direct parallel code \textsc{Dirccr12}
\citep{dirccr12}.


Comparison of the \cc -- He PES with existing surfaces
\citep{akinojo03,topic05} showed an excellent agreement. The \cc
-- para-\hh($J=0$) interaction (obtained by averaging the \cc --
H$_2$ PES over $\theta_2$ and $\phi$) is qualitatively similar to
the \cc -- He PES with a deeper minimum (see values at the end of
present Section). As illustrated in Figure~\ref{fig:PES}, these
PES are largely dominated by the rod-like shape of \cc, implying a
prolate ellipsoid symmetry of the equipotentials.


In a second step, let us consider how to circumvent the difficulty of
the angular expansion of the above PES, in order to obtain reliable
expansions for He and H$_2$ (eqs \ref{eq:pot} and \ref{eq:pot2}).

Using the angular spline representation, we first expressed each
PES over a fine $\theta_1$ mesh suitable for a subsequent high
$l_1$ expansion. As expected from the work of \citet{chapman77},
high $l_1$ expansions (\ref{eq:pot}) resulted in severe Gibbs
oscillations for $R$ in the range 5--7 Bohr, spoiling completely
the description of the low energy features of the PES. Then,
having in mind low energy scattering applications, we regularized
the PES by introducing a scaling function $S_f$. We replaced
$V(R,\theta_1,...)$ by $S_f(V(R,\theta_1,...))$, where $S_f(V)$
returns $V$ when $V$ is lower than a prescribed threshold, and
then smoothly saturates to a limiting value when $V$ grows up into
the repulsive walls. Consequently, the regularized PES  retains
only the low energy content of the original PES, unmodified up to
the range of the threshold energy; it should not be used for
higher collisional energies. However, in contrast to the original
PES, it can be easily expanded over Legendre functions to an
excellent accuracy and is thus suitable for quantum close coupling
studies. We selected a threshold value of 300 cm$^{-1}$, and
improved the quality of the expansion by applying a weighted
fitting strategy \citep[e.g.][]{hodges04} to focus the fit on the
details of the attractive and weakly repulsive regions of the PES.
Using $l_1\leq 35$, both the He and H$_2$ PES fits were converged
to within 1 cm$^{-1}$ for $V \le 300$ cm$^{-1}$. These expansions
still describe the range $300<V<1000$~cm$^{-1}$ to within an
accuracy of a few $\rm cm^{-1}$.

The corresponding absolute minima are the following (in cm$^{-1}$ and
Bohr): for \cc -- He, $V=-40.25$ for $R=6.32$ and
$\theta_1=95.2^{\circ}$; for \cc -- para-\hh ($J=0$), $V=-111.24$ for
$R=6.41$ and $\theta_1=94.0^{\circ}$; and for \cc -- \hh, $V=-192.49$
for $R=9.59$, $\theta_1=180^{\circ}$, and $\theta_2=0^{\circ}$.

\section{Inelastic cross section and rates}\label{sec:cross}

In the following $J_1, J^\prime_1$, denote the initial and final
angular momentum of the \cc molecule, respectively, and $J_2$ denote
the angular momentum of H$_2$. We also denote the largest value of
{$J_1, J^\prime_1$} as $J_{1\rm up}$.

The most reliable approach to compute inelastic cross sections
$\sigma_{J_1J^{\,\prime}_1}(E)$ is to perform quantum close
coupling calculations. In the case of molecules with a small
rotational constant, like \cc \citep[$B=4549.059\mbox{~MHz}$, see
e.g. ][]{thor00}, quantum calculations become soon intractable,
because of the large number of open channels involved. While
observations at cm-mm wavelengths culminates with $J_{1\rm up}
\lesssim 24$ \citep{kahane94}, sub-mm observations can probe
transition as high as $J_{1\rm up} = 40$, at a frequency of
363.785~GHz and a rotational energy of $202.08 \:\rm
cm^{-1}$~\citep{pepe04,charnley04,cauxpc}. It is thus necessary to
compute rates with transitions up to $J_1=50$ ($E=
386.8$~cm$^{-1}$), in order to properly converge radiative
transfer models.  Also, we aim at computing rates up to a
temperature of 100~K for H$_2$. We resorted to two methods in
order to perform this task. For $J_{1\rm up}\leq 15$, we performed
quantum inelastic scattering calculations, as presented in next
subsection~\ref{par:molscat}. For $J_{1\rm up} > 15$, we used the
QCT method, as presented in subsection \ref{par:rates}.

For He, of less astrophysical importance ([He]/[H]$\sim 0.1$), only
quantum calculations were performed and were limited to the low
temperature regime ($T$=5$-$20~K and $J_1<10$).

\subsection{Rotational inelastic cross sections with
 \textsc{Molscat}}\label{par:molscat}

All calculations were made using the rigid rotor approximation, with
rotational constants $B_{\rm HC_3N}=0.151739$ cm$^{-1}$ and $B_{\rm
H_2}=60.853$ cm$^{-1}$, using the \textsc{Molscat} code
\citep{molscat}. All quantum calculations for \cc -- ortho-\hh were
performed with $J_{\rm H_2}\equiv J_2=1$. Calculations for \cc --
para-\hh were performed with $J_2=0$. We checked at $E_{\rm
tot}=E_{\rm coll}+E_{\rm rot} = 30 \,\rm cm^{-1}$ that the inclusion
of the closed $J_2=2$ channel led to negligible effects.

The energy grid was adjusted to reproduce all the details of the
resonances, as they are essential to calculate the rates with high
confidence \citep{dubernet02,dubernet03,wernli06}. The energy grid and
the quantum methods used are detailed in table~\ref{tab:param}. Using
this grid, we calculated the whole resonance structure of all the
transitions up to $J_1=15$ for the \cc -- para-\hh collisions. At
least 10 closed channels were included at each energy to fully
converge the \cc rotational basis. We used the hybrid
log-derivative/Airy propagator \citep{alexander87}. We increased the
parameter \textsc{STEPS} at the lowest energies to constrain the step
length of the integrator below 0.1 to 0.2 Bohr, in order to properly
follow the details of the radial coefficients. Other propagation
parameters were taken as the \textsc{molscat} default values.

\begin{table}
  \caption{Details of the quantum \textsc{Molscat} cross section
    calculation parameters. $J_{1\rm up}({\rm HC_3N}) \leq 15$ (10 for
    He). Methods: CC, Close coupling, CS, Coupled states approx.,
    IOS, Infinite Order Sudden approx.}
  \label{tab:param} \centering
  \begin{tabular}{ccc}
    \hline\hline \multicolumn{3}{c}{$ \sigma_{J_1J'_1}(E)$, \cc --
      para-\hh collisions}\\ \hline
    $E_{\rm tot}\:\rm (cm^{-1})$ & Energy step ($\rm cm^{-1}$) & Method
    \\
    \hline
    $0.3 \rightarrow 60$ & $0.1$ & CC \\
    $60 \rightarrow 110$ & $10$ & CC \\
    $40 \rightarrow 200$ & $10$ & CS \\
    $50 \rightarrow 800$ & $10-100 $ & IOS\\
    \hline
    \multicolumn{3}{c}{$ \sigma_{J_1J'_1}(E)$, \cc -- ortho-\hh collisions}\\
    \hline $0\rightarrow 30 $& $1$ & CC \\
    \hline
    \multicolumn{3}{c}{$ \sigma_{J_1J'_1}(E)$, \cc -- He collisions}\\
    \hline $0\rightarrow 25 $& $0.1$ & CC \\
    \hline $25\rightarrow 100 $& $5$ & CC \\
    \hline $100\rightarrow 150 $& $10$ & CC \\
    \hline\hline
  \end{tabular}
\end{table}

\begin{figure}
   \resizebox{\hsize}{!}{\includegraphics{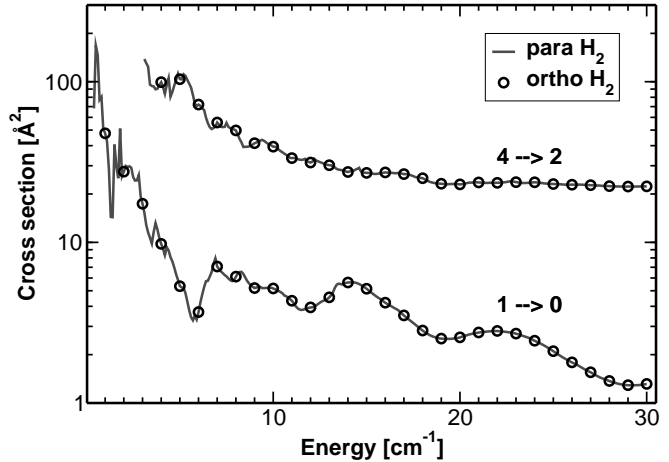}}
 \caption{ \cc -- \hh collisions. Quantum deexcitation cross sections
for transitions $J_1=1\rightarrow 0$ (lower trace) and $J_1=4
\rightarrow 2$ (upper trace) as a function of the energy
$E$=$E_{\rm coll}+E_{\rm rot(HC_3N)}$. Full line, para-\hh ($J=0$)
collisions (0.1 cm$^{-1}$ energy spacing); open circles, ortho-\hh
$(J=1)$. The fine energy grid emphasizes the resonances for $E
\lesssim 10\,\rm cm^{-1}$.}
\label{fig:sectionop}
\end{figure}

Two examples of deexcitation cross-sections are shown in
figure~\ref{fig:sectionop}. We see that for energies between
threshold and about 20 cm$^{-1}$ above threshold, the
cross-section displays many shape resonances, justifying \textit{a
posteriori} our very fine energy grid. This behaviour is by no
means unexpected and is very similar to most earlier calculations
is many different systems, see e.g. \citet{dubernet02,wernli06}
for a discussion. In a semi-classical point of view, those shape
resonances manifest the trapping of the wave-packet between the
inner repulsive wall and the outer centrifugal barrier, see
\citet{wie03,abrol01}. At energies higher than about 20 cm$^{-1}$
above threshold, all cross-sections become smooth functions of the
energy.

Figure~\ref{fig:sectionop} also shows that ortho-\hh inelastic
cross-sections follow very closely the para-\hh ones, including the
position of resonances. Examination of all cross-sections reveals that
the relative difference between $\sigma_{J_1J'_1}(E, \mbox{para})$ and
$\sigma_{J_1J'_1}(E, \mbox{ortho})$ is less than $5\%$. This justifies
\textit{a posteriori} the much smaller amount of computational effort
devoted to ortho-\hh collisions as well as the neglect of $J_2 = 2$
closed para-\hh channels. A detailed discussion of this behaviour is
put forward in section~\ref{sec:paraortho}.

\subsection{Quantum rates and classical rates}\label{par:rates}

The quantum collisional rates are calculated for $J_{1\rm up}\leq
15$, at astrophysically relevant temperatures, from 5~K to 100~K.
We average the cross-sections described in the preceding section
over the Maxwell distribution of velocities, up to a kinetic
energy at least 10 times $kT$. The quantum calculations at the
higher end of the energy range are approximated at the IOS level
(see table \ref{tab:param}), which is justified at these energies
by the smallness of the rotational constant $B_{\rm HC_3N}$. Also
we used a coarse energy grid for the IOS calculations because the
energy dependence of the cross-sections becomes very smooth.

For values of $J_1 > 15$, the close coupling approach enters a
complexity barrier due to the rapid increase of the number of channels
involved in calculations, while memory and CPU requirements scale as
the square and the cube of this number, respectively. Resorting to
quantum CS or IOS approximations is inaccurate, because the energy is
close to threshold for high-$J_1$ channels. In the meanwhile, the
accuracy of classical approximations improves for higher collisional
energies. For the energy range where $J_1 > 15$ channels are open
\emph{and} for deexcitation processes involving those channels, we
employ a Quasi-Classical Trajectory (QCT) method, which has been shown
at several instances to be a valid approximation for higher
collisional energies and large rates~\citep{chapman77,lepp95,mandy04,
faure06}.

For Monte-Carlo QCT methods, we must devise a way of defining an
ensemble of initial conditions for classical trajectories, on the
one hand, and of analyzing the final state of each trajectory, on
the other hand. Contrary to the asymmetric rotor case \citep[like
water, see][]{faure04}, the analysis of final conditions for a
linear molecule is straightforward. Using the simplest
quantization approximation, we bin the final classical angular
momentum $J'_1$ of \cc to the nearest integer. While the quantum
formalism goes through a microcanonical calculation ---calculating
$\sigma_{J_1J_1'}(E)$ for fixed energies, then averaging over
velocity distributions--- it is possible for QCT calculations to
directly resort to a canonical formalism, i.e. to select the
initial velocities of the Monte-Carlo ensemble according to the
relevant Maxwell-Boltzmann distribution and find the rates  as:
\begin{equation}\label{eq:rate}
k_{J_1J'_1} = \left(\frac{8kT}{\pi\mu}\right)^{1/2}\,\pi
b_{max}^2\,\frac{N}{N_{\rm tot}}
\end{equation}
where $b_{max}$ is the maximum impact parameter used (with the impact
parameter $b$ distributed with the relevant $b\,\textmd{d}b$
probability density) and $N$ is the number of trajectories with the
right final $J'_1$ value among all $N_{\rm tot}$ trajectories. The
Monte-Carlo standard deviation is:
\begin{equation}\label{eq:error}
    \frac{\delta k_{J_1J_1'}}{k_{J_1J_1'}} = \left(\frac{N_{\rm
    tot}-N}{N_{\rm tot}N}\right)^{1/2} \quad ,
\end{equation}
showing that the accuracy of the method improves for larger rates.
The $b_{max}$ parameter was determined by sending small batches of
$500$ to $1,000$ trajectories for fixed $b$ values; values of $20 \leq
b_{max} \leq 26$ Bohr were found. We then sent batches of $10,000$
trajectories for each temperature in the range $5-100$K, with a step
of 5K. Trajectories are integrated by means of a B\"{u}rlich-Stoer
algorithm~\citep{numrec92}, with a code similar to that of
\citet{faure05a}. Precision is checked by conservation of total energy
and total angular momentum.

\begin{figure}\centering
 \resizebox{\hsize}{!}{\includegraphics[angle=-90]{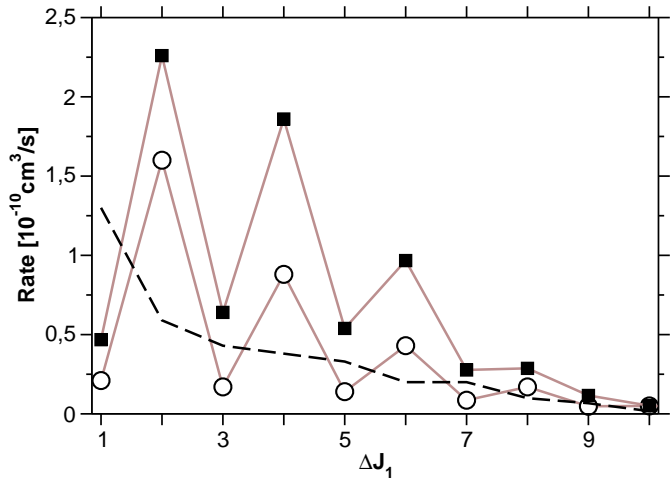}}
 \caption{ Collisional excitation rates for \cc, from $J_{1}=0$, at 20~K.
Present quantum close-coupling rates: Open circles, \cc -- He
collisions; filled squares, \cc -- para-\hh~($J=0$). Gray lines
serve as a guide to the eyes. For sake of comparison, QCT rates
for the \cc -- He \protect\citep{green78} are shown as dashed
line.}
  \label{fig:ratecompare}
\end{figure}

\begin{figure*}\centering
  \begin{tabular}{lr}
    \includegraphics[width=8cm]{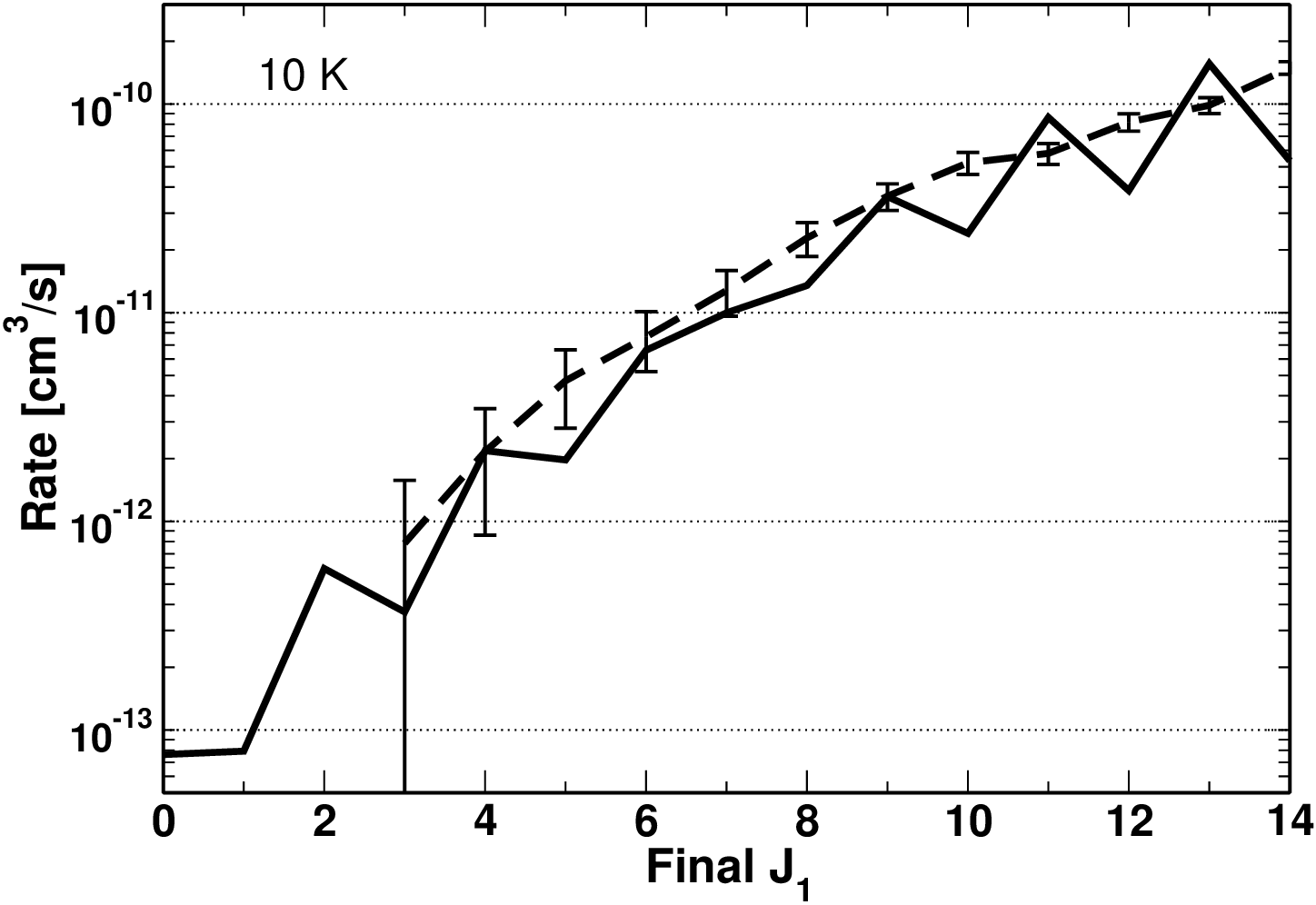} &
    \includegraphics[width=8cm]{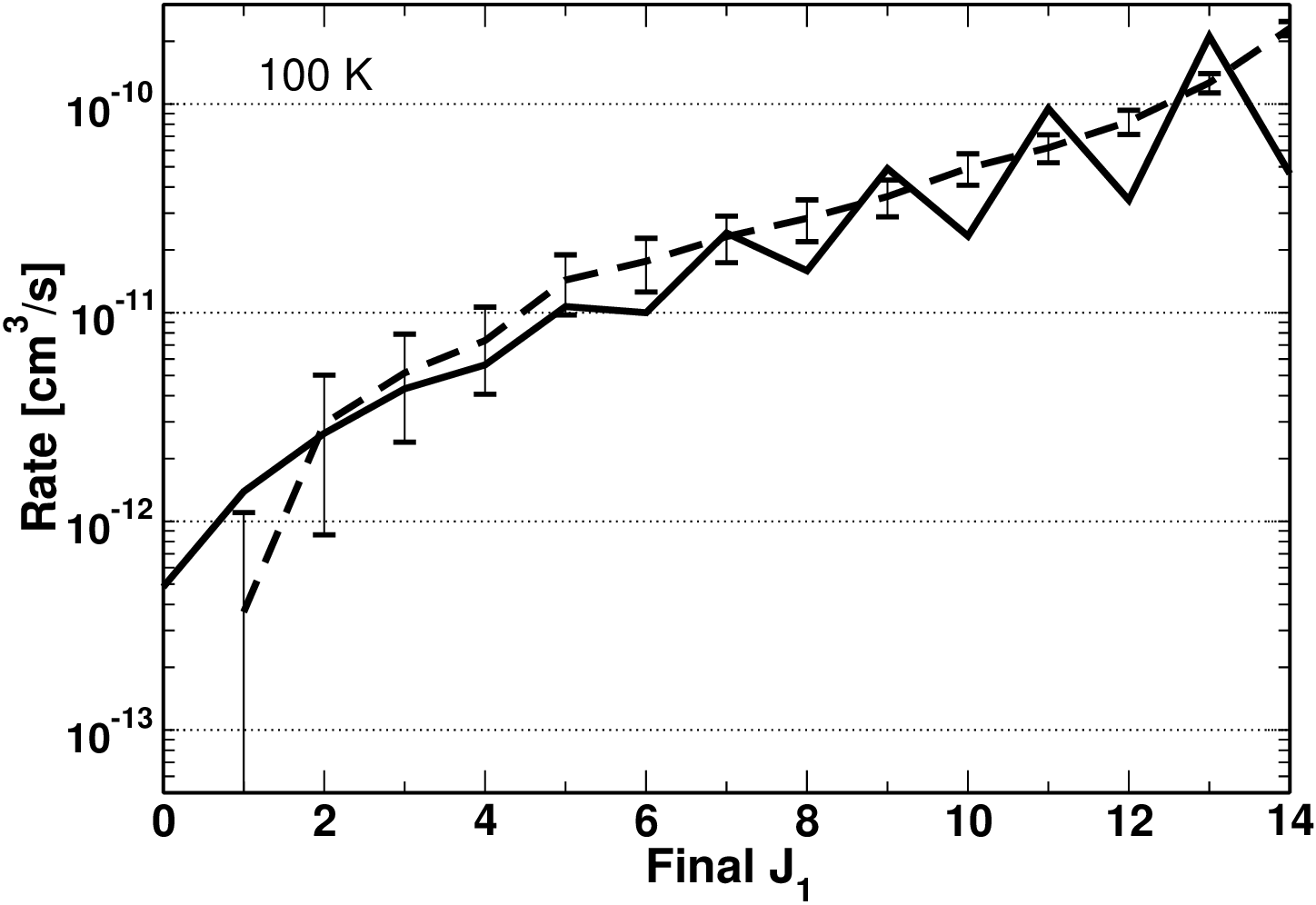}
  \end{tabular}
  \caption{\cc -para \hh collisions. Deexcitation rates, from
      $J_{1}=15$, at 10~K (left panel) and 100~K (right
      panel).  Full black line, quantum calculations; dashed line,
      Monte-Carlo quasi-classical calculations with error bars.}
      \label{fig:rate12}
\end{figure*}

Some illustrative results are shown in tables~\ref{tab:rates} and
~\ref{tab:rates2} and are illustrated in figures~\ref{fig:ratecompare}
and~\ref{fig:rate12}.

As an alternative to QCT calculations, we tested J-extrapolation
techniques, using the form of \citet{depristo79} generally used by
astrophysicists (see for example \citet{lamda}, section 6). We
found that even if it reproduces the interference pattern, the
extrapolation systematically underestimates the rates, for $J_1\ge
20$. Hence, QCT rates are more precise in the average.

\begin{table}
  \centering \caption{\cc -- para-\hh~s($J=0$)collisions. Quantum deexcitation
rates in $\mathrm{cm^3\,s^{-1}}$, for $J_1'=0$, for successive
initial $J_1$ and for various temperatures. Powers of ten are
denoted in parenthesis.} \label{tab:rates}
\begin{tabular}{ccccc} \hline\hline
&\\
 $J_{1}$  & $T = 10 \rm K\;$ &  $T = 20 \rm K\;$ & $T = 50 \rm
K\;$ & $T = 100 \rm K\;$ \\
&\\
 \hline
&\\
1  &  2.03(-11)  &  1.59(-11)  &  1.32(-11)  &  1.24(-11) \\
2  &  4.94(-11)  &  4.83(-11)  &  6.23(-11)  &  8.04(-11) \\
3  &  1.20(-11)  &  1.04(-11)  &  8.23(-12)  &  7.43(-12) \\
4  &  2.25(-11)  &  2.57(-11)  &  2.85(-11)  &  2.87(-11) \\
5  &  7.01(-12)  &  6.80(-12)  &  5.62(-12)  &  4.77(-12) \\
6  &  9.15(-12)  &  1.18(-11)  &  1.42(-11)  &  1.38(-11) \\
7  &  3.14(-12)  &  3.40(-12)  &  3.46(-12)  &  3.26(-12) \\
8  &  2.45(-12)  &  3.71(-12)  &  5.92(-12)  &  6.61(-12) \\
9  &  1.63(-12)  &  1.63(-12)  &  1.95(-12)  &  2.18(-12) \\
10  &  5.35(-13)  &  8.13(-13)  &  2.00(-12)  &  2.96(-12) \\
11  &  7.81(-13)  &  7.01(-13)  &  9.42(-13)  &  1.36(-12) \\
12  &  1.37(-13)  &  1.58(-13)  &  6.17(-13)  &  1.32(-12) \\
13  &  2.74(-13)  &  2.51(-13)  &  4.17(-13)  &  8.26(-13) \\
14  &  4.14(-14)  &  4.65(-14)  &  2.24(-13)  &  6.28(-13) \\
15  &  7.63(-14)  &  8.26(-14)  &  1.76(-13)  &  4.85(-13) \\
\hline\hline
\end{tabular}
\end{table}

\begin{table}
\centering \caption{\cc -- para-\hh collisions. Quantum or QCT
($^\dag$) deexcitation rates in $\mathrm{cm^3\,s^{-1}}$, for
$J_{1}-J'_{1}=1,2,3,4$, for various temperatures and
representative values of $J_{1}$. Powers of ten are denoted in
parenthesis.} \label{tab:rates2}
\begin{tabular}{llcccc} \hline\hline
&\\ $J_{1}'$ &$J_{1}$& $T = 10 \rm K\;$ & $T = 20 \rm K\;$ &
 $T = 50 \rm K\;$ & $T = 100 \rm K\;$ \\
&\\
 \hline
&\\
0  &  1  &  2.03(-11) & 1.59(-11)   &  1.32(-11)  &  1.24(-11) \\
0  &  2  &  4.94(-11) &  4.83(-11)  &  6.23(-11)  &  8.04(-11) \\
0  &  3  &  1.20(-11) &  1.04(-11)  &  8.23(-12)  &  7.43(-12) \\
0  &  4  &  2.25(-11) &  2.57(-11)  &  2.85(-11)  &  2.87(-11) \\
\\
5  &  6  &  6.34(-11)  &  5.48(-11)  &  4.80(-11)  &  4.64(-11) \\
5  &  7  &  1.30(-10)  &  1.38(-10)  &  1.72(-10)  &  2.04(-10) \\
5  &  8  &  3.93(-11)  &  3.66(-11)  &  3.27(-11)  &  3.21(-11) \\
5  &  9  &  6.83(-11)  &  7.61(-11)  &  8.63(-11)  &  8.93(-11) \\
\\
10  &  11  &  5.77(-11)  &  5.35(-11)  &  4.75(-11)  &  4.61(-11) \\
10  &  12  &  1.50(-10)  &  1.53(-10)  &  1.81(-10)  &  2.11(-10) \\
10  &  13  &  3.91(-11)  &  3.80(-11)  &  3.50(-11)  &  3.40(-11) \\
10  &  14  &  8.51(-11)  &  8.84(-11)  &  9.42(-11)  &  9.47(-11) \\
\\
\hline
\\
15  &  16 $^\dag$  &  1.45(-10) &  1.49(-10) &  1.80(-10) &  2.28(-10) \\
15  &  17 $^\dag$  &  1.06(-10) &  1.03(-10) &  9.60(-11) &  1.18(-10) \\
15  &  18 $^\dag$  &  8.71(-11) &  9.38(-11) &  7.93(-11) &  8.24(-11) \\
15  &  19 $^\dag$  &  7.59(-11) &  6.81(-11) &  6.23(-11) &  7.10(-11) \\
\\
25  &  26 $^\dag$  &  1.14(-10) &  1.50(-10) &  1.83(-10) &  2.30(-10) \\
25  &  27 $^\dag$  &  1.13(-10) &  1.05(-10) &  1.18(-10) &  1.31(-10) \\
25  &  28 $^\dag$  &  8.55(-11) &  8.38(-11) &  8.34(-11) &  7.76(-11) \\
25  &  29 $^\dag$  &  7.67(-11) &  7.45(-11) &  8.31(-11) &  7.02(-11) \\
\\
35  &  36 $^\dag$  &  1.16(-10) &  1.34(-10) &  1.73(-10) &  2.32(-10) \\
35  &  37 $^\dag$  &  9.63(-11) &  1.12(-10) &  1.21(-10) &  1.11(-10) \\
35  &  38 $^\dag$  &  8.33(-11) &  9.20(-11) &  8.56(-11) &  9.19(-11) \\
35  &  39 $^\dag$  &  8.77(-11) &  8.51(-11) &  7.65(-11) &  7.51(-11) \\
\\
\hline\hline
\end{tabular}
\end{table}

For H$_2$, all deexcitation rates $k_{J_1J_1'}(T)$, $J_1\neq J_1'\leq
50$, are fitted with the following formula \citep{wernli06}:
\begin{equation}\label{eq:fit}
    \log_{10}\left(k_{J_1J_1'}(T)\right)=\sum_{n=0}^{
    4}a^{(n)}_{J_1J_1'} x^n
\end{equation}
where $x=T^{-1/6}$. As some transitions have zero probability
within the QCT approach, the above formula was employed when rates
were bigger than 10$^{-12}$ cm$^3$s$^{-1}$ for at least one grid
temperature. For these rates, null grid values were replaced by a
very small value, namely 10$^{-14}$ cm$^3$s$^{-1}$, to avoid fitting
irregularities. All rates not fulfilling this condition are set to
zero. Note that below 20~K, QCT rates for low-probability
transitions may show a non physical behaviour. All
$a^{(n)}_{J_1J_1'}$ coefficients are provided as online material,
for a temperature range $5{\rm\; K}\leq T \leq 100{\rm\; K}$. We
advise to use the same rates for collisions with ortho-\hh as for
para-H$_2$, since their difference is smaller than the uncertainty
on the rates themselves. Rates with He were not fitted, but can be
obtained upon request to the authors.

\section{Discussion}\label{sec:disc}

\subsection{Para and ortho \hh
cross-sections}\label{sec:paraortho} A comparison of the
$\sigma_{J_1J_1'}(E)$ cross sections for \cc with ortho-\hh and
para-\hh is given in figure~\ref{fig:sectionop}. It can be seen that
the difference between the two spin species of \hh may be considered
as very small, in any case smaller than other PES and cross-section
uncertainties. This is an unexpected result, as sizeable differences
between para-\hh and ortho-\hh inelastic cross-sections exist for
other molecules. These differences were expected to increase for a
molecule possessing a large dipolar moment, in view of the results
obtained for the C$_2$ molecule \citep{phillips94}, the CO molecule
\citep{wernli06}, the OH radical \citep{offer94}, the NH$_3$ molecule
\citep{offer89,flower94} and the $\rm H_2O$ molecule
\citep{phillips96,dubernet02,dubernet03,dubernet06}, due to the
interaction between the dipole of the molecule and the quadrupole of
\hh(J$_2$ $>$ 0).

This apparently null result deserves an explanation. We focus on
equation~(9) of \citet{green75}. This equation describes the different
matrix elements that couple the various channels in the close-coupling
equations. Some triangle rules apply which restrict the number of
terms in the sum of equation (9); the relevant angular coupling
algebra is represented there as a sum of terms of the type
\begin{equation}\label{eq:green75}
    \left(\begin{array}[c]{ccc} l &L'& L \\ 0 & 0 & 0
    \end{array}\right)
    \;
 \left(\begin{array}[c]{ccc} l_1 &J'_1& J_1 \\ 0 & 0 & 0
    \end{array}\right)
    \;
     \left(\begin{array}[c]{ccc} l_2 &J'_2& J_2 \\ 0 & 0 & 0
    \end{array}\right)
    \;
     \left\{\begin{array}[c]{ccc} L'& L & l \\ J_{12} & J'_{12} &
     J
    \end{array}\right\}\quad ,
\end{equation}
where we have the potential function expanded in terms of Eqs. (4) and
(A2) in \citet{green75}, by means of the coefficients $v_{l_1l_2
l}$. The symbol $(\ldots)$ are 3-$j$ symbols, the $\{\ldots\}$ is a
6-$j$ symbol, see \citet{messiah69}. We also define
$\vec{J}_{12}=\vec{J}_1+\vec{J}_2$. We have the following rules:
\begin{itemize}
\item The para-\hh inelastic collisions are dominated by the $J_2 = 0$
channel (the $J_2=2$ channel is closed till $E_{coll} \gtrsim 365.12
\;\rm cm^{-1}$). Then, only the $l_2=0$ may be retained
($J_2=J'_2=0$), due to the third 3-$j$ symbol in
eq.(\ref{eq:green75}).
\item The ortho-\hh remains in $J_2=1$, implying $l_2=0,2$.

\item For inelastic collisions, $J_1\neq J'_1$ implies potential terms
with $l\neq 0$, because of the 6-$j$ term in
Eq.(\ref{eq:green75}). Indeed, $J_2=J'_2$ and $J_1\neq J'_1$ entail
$J_{12}\neq J'_{12}$.
\end{itemize}

\begin{figure}
  \resizebox{\hsize}{!}{\includegraphics{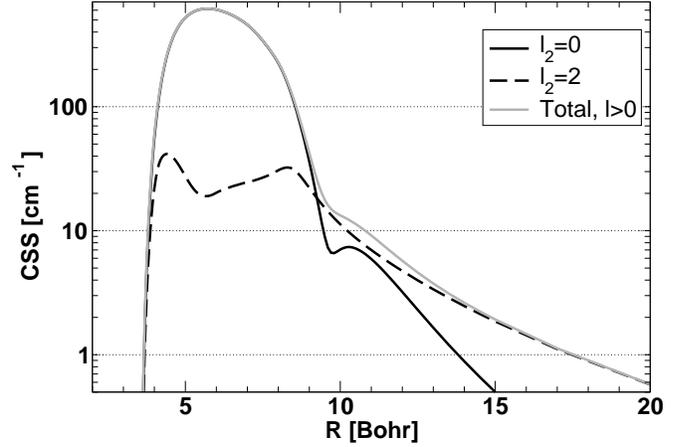}}
  \caption{Comparison of the coupling terms ($l\ne 0$) in the \cc --
    \hh potential as a function of the intermolecular distance. CSS is
    $ \left(\sum_{l\ne 0, l_1} v_{l_1 l_2 l}(R)^2\right)^{1/2}$. Terms
    with $l_2=0$ are common to ortho and para \hh, while the $l_2=2$
    curve represents purely ortho terms. See text.}
  \label{fig:comp}
\end{figure}

The key point is thus to compare the $v_{l_1l_2 l}(R)$ coefficients
(eq. \ref{eq:pot2}) with $l \neq 0$ in the two cases:
\begin{itemize}
  \item{$l_2=0$} para and ortho contributions;
  \item{$l_2=2$} ortho contribution only.
\end{itemize}
Figure \ref{fig:comp} displays such a comparison. We notice that
the coupling is largely dominated by the $l_2=0$ contribution,
terms which are common to collisions with para and ortho
conformations. This is particularly true for $R<10$ Bohr, the
relevant part of the interaction for collisions at temperatures
higher than a few Kelvin. At a higher intermolecular separation,
terms implied only in collisions with ortho-\hh become dominant,
but in this regime the potential is also less than a few
cm$^{-1}$. Sizeable differences in rates between ortho and para
forms are thus expected only either at very low temperatures, or
possibly at much higher temperatures, with the opening of
\hh(J$_2=2,3$) channels.

\subsection{Propensity rules}
In figure \ref{fig:ratecompare}, we compare the various rates that
we obtain here with the ones previously published by
\citet{green78}. These authors used a coarse electron-gas
approximation for the PES, and computed rates by a QCT classical
approach. Despite these approximations, we see that the rates
obtained by \citet{green78} are qualitatively comparable with the
quantum rates obtained here, in an average way. However, as table
\ref{tab:rates2} and figures \ref{fig:rate12} and
\ref{fig:ratecompare} show clearly, only quantum calculations
manifest the strong $\Delta J = 2$ propensity rule. This rule
originates in the shape of the PES, being nearly a prolate
ellipsoid, dominated by the rod shape of \cc and \emph{not
dominated} by the large dipole of HC$_3$N molecule (3.724~Debye).
Because of the very good approximate symmetry
$\theta_1\leftrightarrow \pi - \theta_1$, the $l_1$ even terms
(equation (\ref{eq:green75}) and \citet{green75}) are the most
important ones, directing the inelastic transition toward even
$\Delta J_1$. This propensity has also been explained
semi-classically by \citet{miller77} in terms of an interference
effect related to the even anisotropy of the PES. These authors
show in particular that the reverse propensity can also occur if
the odd anisotropy of the PES is sufficiently large. This reverse
effect is indeed observed in Fig.~\ref{fig:rate12} for transitions
with $\Delta J>10$. A similar propensity rule has been
experimentally observed for CO--He collisions \citep{sims04}.

Besides this strong $\Delta J = 2$ propensity rule, one can see
from table~\ref{tab:rates2} and figures \ref{fig:ratecompare},
\ref{fig:rate12} that the rod-like interaction drives large
$\Delta J$ transfers. For instance, for T $> 20$~K, rates for
$\Delta J > 6$ are generally larger than rates for $\Delta J = 1$,
and rates for $\Delta J > 8$ are only one order of magnitude below
those for $\Delta J = 2$. This behaviour is likely to emphasize
the role of collisional effects versus radiative ones. This
effect, of purely geometric origin, has been predicted previously
\citep{bosanac80} and is of even greater importance for longer
rods like $\rm HC_5N$, $\rm HC_7N$, $\rm HC_9N$, see
\citet{snell81,dickinson82}.

We also observe that the ratio
$k_{J_1J'_1}(\textrm{He})/k_{J_1J'_1}(\mbox{para-H$_2$}) $ is in
average close to $1/1.4 \sim 1/\sqrt{2}$, thus confirming the
similarity of He and para-\hh as projectiles, as generally
assumed. But it is also far from being a constant, as already
observed for H$_2$O \citep{phillips96} or CO \citep{wernli06}. Our
data shows that the $1/\sqrt{2}$ scaling rule results in errors up
to 50\%.

\subsection{Population inversion and critical densities}
\begin{figure}
  \resizebox{\hsize}{!}{\includegraphics{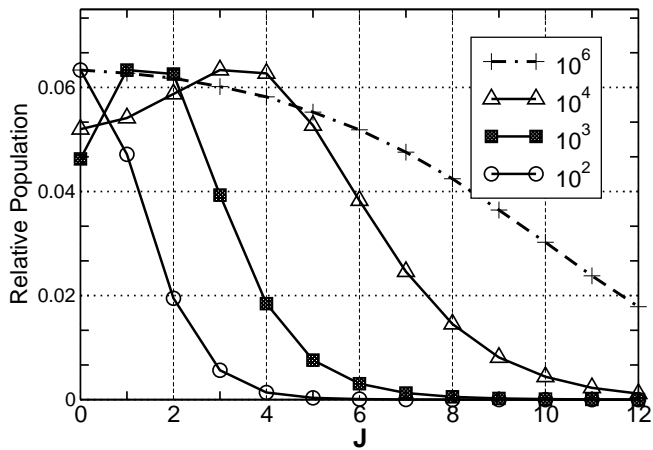}}
  \caption{Population per $J_1,m_{J_1}$-sub-level, following
    equation~(\protect\ref{eq:ss}), for varying \hh densities (in $\rm
    cm^{-3}$) at a kinetic temperature of 40K.}
  \label{fig:invers}
\end{figure}

\begin{figure}
  \resizebox{\hsize}{!}{\includegraphics{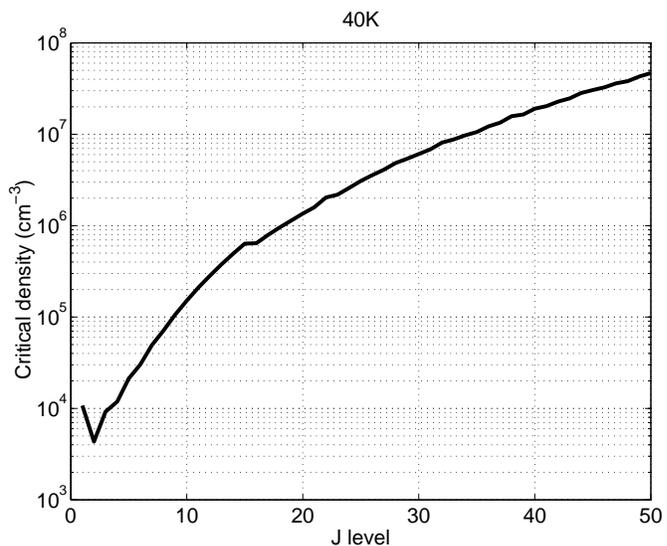}}
  \caption{\hh critical density $n^\star$ (in cm$^{-3}$) for the \cc
    -- para-\hh collisions, at $T =40\,\rm K$, equation
    (\protect\ref{eq:nstar}). For $j\leq 15$, quantum rates, for
    $j>15$, classical rates. The change of method explains the small
    discontinuity between $j=15,16$. The increase of critical density
    at $j=1$ is related to the propensity rule, see figure
    \protect\ref{fig:rate12}.}
  \label{fig:critical}
\end{figure}

Because of the strong $\Delta J_1=0,2,4$ propensity rule,
population inversion could be strengthened if LTE conditions are
not met, even  neglecting hyperfine effects\footnote{Hyperfine
effects in \cc inelastic collisions will be dealt with in a
forthcoming paper, \citet{wie06}} \citep{hunt99}. In order to see
the density conditions giving rise to population inversion, we
solved the steady-state equations for the population of the
$J=0,1,\dots,15$ levels of \cc, including collisions with \hh
(densities ranging from $10^2$ to $10^6 \rm\; cm^{-3}$), a
black-body photon bath at 2.7~K, in the optically thin
approximation, \citep{goldsmith72}~:
\begin{eqnarray}\label{eq:ss}
    \frac{\textmd{d}n_i}{\textmd{d}t}=0&=&
    +\sum_{j\neq i}n_j\,\left[ A_{ji} +
    B_{ji} \;n_\gamma\left(\nu_{ji}\right)+k_{ji}\;
    n_{\rm H_2}\right] \nonumber \\
    & & - n_i\,\sum_{j\neq i}
    \left[A_{ij}+B_{ij}\,n_\gamma\left(\nu_{ij}\right)+k_{ij}\; n_{\rm
    H_2}\right]
\end{eqnarray}
where $i,j$ are the levels, $n_\gamma$ is the photon density at
temperature $T_\gamma$ and $n_{\rm H_2} $ is the hydrogen density
at kinetic temperature $T_{\rm H_2}$. Figure~\ref{fig:invers}
shows the results at $T_{\rm H_2} = 40$~K. The lines show the
population per sub-levels $\left|J_1, m_{J_1}\right>$. For a
consequent range of \hh densities, $10^4\lesssim n_{\rm H_2}
\lesssim 10^6$, population inversion does occur, for $0\leq
J_1\leq 2, 3, 4$. Our new rates are  expected to improve the
interpretation of the lowest-lying lines of \cc, especially so in
the 9 -~20~GHz regions (cm-mm waves), see for example
\citet{walms86,takano98,hunt99}, and \citet{kal04} for a recent
study. Moreover, from the knowledge of both collision coefficients
$k_{ij}$ and Einstein coefficients $A_{ij}$, it is possible to
derive a critical density of \hh, defined as:
\begin{equation}\label{eq:nstar}
    n^{\star}_i(T)=\frac{\sum_{j<i}A_{ij}}{\sum_{j<i}k_{ij}}
\end{equation}
The $n^\star$ density is the \hh density at which photon deexcitation
and collisional deexcitation are equal. The evolution of $n^\star$
with $J_1$ at $T= 40\;\rm K$ is given in figure~\ref{fig:critical}. It
can be seen that for many common interstellar media, the LTE
conditions are not fully met.

It must be underlined that similar effects should appear for the whole
cyanopolyyne ($\rm HC_{5,7,9}N$) family, where cross-sections should
scale approximately with the rod length \citep{dickinson82}. It is
expected that the propensity rule $\Delta J_1=2,4,\dots$ should remain
valid. Also, the critical density should decrease for the higher
members of the cyanopolyyne family, as the Einstein A$_{ij}$
coefficients, hence facilitating the LTE conditions.


\section{Conclusion}

We have computed two {\it ab initio} surfaces, for the \cc -- He
and \cc -- \hh systems. The latter was built using a carefully
selected set of \hh orientations, limiting the computational
effort to approximately five times the \cc -- He one. Both
surfaces were successfully expanded on a rotational basis suitable
for quantum calculations using a smooth regularization of the
potentials. This approach circumvented the severe convergence
problems already noticed by \citet{chapman77} for such large
molecules.  The final accuracy of both PES is a few cm$^{-1}$ for
potential energy below 1000~cm$^{-1}$.

Rates for rotational excitation of \cc by collisions with He atoms
and \hh molecules were computed for kinetic temperatures in the
range 5 to 20~K and 5 to 100~K, respectively, combining quantum
close coupling and quasi-classical calculations. The rod-like
symmetry of the PES strongly favours even $\Delta J_1$ transfers
and efficiently drives large $\Delta J_1$ transfers. Quasi
classical calculations are in excellent agreement with close
coupling quantum calculations but do not account for the even
$\Delta J_1$ interferences.  For He, results compare fairly with
\citet{green78} QCT rates, indicating a weak dependance to the
details of the PES.  For para-H$_2$, rates are compatible in
average with the generally assumed $\sqrt{2}$ scaling rule, with a
spread of about 50 \%. Despite the large dipole moment of
$\mathrm{HC_3N}$, rates involving ortho-H$_2$ are very similar to
those involving para-H$_2$, due to the predominance of the rod
interactions.

A simple steady-state population model shows population inversions for
the lowest \cc levels at \hh densities in the range
10$^4-$10$^6$~cm$^{-3}$. This inversion pattern manifests the
importance of large angular momentum transfer, and is enhanced by the
even $\Delta J_1$ quantum propensity rule.

The \cc molecule is large enough to present an original
collisional behaviour, where steric hindrance effects hide the
details of the interaction, and where quasi classical rate
calculations achieve a fair accuracy even at low temperatures.
With these findings, approximate studies for large and heavy
molecules should become feasible including possibly the modelling
of large $\Delta J$ transfer collisions and ro-vibrational
excitation of low energy bending or floppy modes.


\begin{acknowledgements}
This research was supported by the CNRS national program
``Physique et Chimie du Milieu Interstellaire'' and the ``Centre
National d'Etudes Spatiales''.  LW was partly supported by a
CNRS/NSF contract. MW was supported by the Minist\`ere de
l'Enseignement Sup\'erieur et de la Recherche. CCSD(T)
calculations were performed on the IDRIS and CINES French national
computing centers (projects no. 051141 and x2005 04 20820).
\textsc{Molscat} and QCT calculations were performed on local
workstations and on the ``Service Commun de Calcul Intensif de
l'Observatoire de Grenoble'' (SCCI) with the valuable help from F.
Roch.
\end{acknowledgements}


\bibliographystyle{aa}
\bibliography{cyano_v3}

\appendix

\newcounter{onltable}
\renewcommand\thetable{A\theonltable}
\addtocounter{onltable}{1}

\begin{longtable}{c c c c c c c}
  \caption{Fitting coefficients of HC$_3$N-para-H$_2$($J_2$=0) rates,
      following formula (\ref{eq:fit}). $J_1$ and $J'_1$ are the initial and
      final rotation states, respectively. The rates thus obtained are
      in $\rm cm^3s^{-1}$. Rates are quantum for $J_{1 \rm up} \le 15$,
      classical otherwise } \\

  \input{coeffs_CC-clas.tex}
\end{longtable}

\end{document}

%% file: coeffs_CC-clas.tex
$J_1$ & $J'_1$ & $a^{(0)}$ & $a^{(1)}$ & $a^{(2)}$ & $a^{(3)}$ & $a^{(4)}$  \\ \\
\hline
\hline
   1 & 0 &    -25.51832 &    106.88002 &   -291.58814 &    350.10316 &   -154.84199\\
   2 & 0 &    -24.35512 &    106.84051 &   -287.22039 &    328.93070 &   -136.65661\\
   2 & 1 &    -14.61936 &     33.10754 &   -100.86405 &    134.46313 &    -64.92410\\
\hline
   3 & 0 &     -5.47571 &    -35.76987 &     79.22847 &    -71.33148 &     21.84669\\
   3 & 1 &    -14.72047 &     37.31354 &    -99.10777 &    109.36028 &    -43.08167\\